\documentclass[aps,pra,superscriptaddress,amsmath,amssymb,preprintnumbers]{revtex4}

\usepackage{amssymb}
\usepackage{booktabs}
\usepackage{graphicx}   
\usepackage{bm}
\usepackage{color}
\usepackage{cancel}
\usepackage{slashed}
\usepackage{soul}

\begin{document}

\title{Bounds on mixed state entanglement}

\newcommand{\orcidauthorA}{0000-0000-000-000X} 

\author{Bruno Leggio $^{1}$, Anna Napoli $^{2,4}$, Hiromichi Nakazato  $^{5}$ and Antonino Messina $^{3,4}$}


\affiliation{$^{1}$ Laboratoire Reproduction et D{\'e}veloppement des Plantes, Univ Lyon, ENS de Lyon, UCB Lyon 1, CNRS, INRA, Inria, Lyon 69342, France\\
$^{2}$  Universit\`a degli Studi di Palermo, Dipartimento di Fisica e Chimica - Emilio Segr\`e, Via Archirafi 36, I-90123 Palermo, Italy\\
$^{3}$   Universit\`a degli Studi di Palermo, Dipartimento di Matematica ed Informatica, Via Archirafi, 34, I-90123 Palermo, Italy \\
$^{4}$  I.N.F.N. Sezione di Catania, Via Santa Sofia 64, I-95123 Catania, Italy\\
$^{5}$  Department of Physics, Waseda University, Tokyo 169-8555, Japan\\
 }






\begin{abstract}
In the general framework of $d\times d$ mixed states, we derive an explicit bound for bipartite NPT entanglement based on the mixedness characterization of the physical system. The result derived is very general, being based only on the assumption of finite  dimensionality. In addition it turns out to be of experimental interest since some purity measuring protocols are known. Exploiting the bound in the particular case of thermal entanglement, a way to connect thermodynamic features to monogamy of quantum correlations is suggested, and some recent results on the subject are given a physically clear explanation.
\end{abstract}

\maketitle

\section{Introduction}
In dealing with mixed states of a physical system, one has to be careful when speaking about entanglement. The definition of bipartite mixed state entanglement is unique (although problems may arise in dealing with multipartite entanglement \cite{Guhne}), but its quantification relies on several different criteria and it is not yet fully developed: many difficulties arise already in the definition of physically sensible measures \cite{Horodecki,Mintert}. The main problem affecting a few known mixed state entanglement measures is, indeed, the fact that extending a measure from a pure state case to a mixed state case usually requires challenging maximization procedures over all its possible pure state decompositions \cite{Eisert}-\cite{Takayanagi}. Notwithstanding, the investigation of the connection between entanglement and mixedness exhibited by a quantum system is of great interest, e.g. in quantum computation theory \cite{Jozsa,Datta} as well in quantum teleportation \cite{Paulson}. The threshold of mixedness exhibited by a quantum system compatible with the occurrence of entanglement between two parties of the same system has been analyzed, leading for example to the so-called Kus-Zyczkowski ball of absolutely separable states \cite{Kus}-\cite{Aubrun}. Quite recently, possible links between entanglement and easily measurable observables have been exploited to define experimental protocols aimed at measuring quantum correlations \cite{Brida}-\cite{Johnston}. The use of measurable quantities as  entanglement witnesses for a wide class of systems has been known for some time \cite{Toth,Wiesniak}, but an analogous possibility amounting at entanglement measures is a recent and growing challenge. To present day some bounds for entanglement are measured in terms of correlation functions in spin systems \cite{Cramer} or using quantum quenches \cite{Cardy}. Indeed an experimental measure of entanglement is, generally speaking, out of reach because of the difficulty in addressing local properties of many-particle systems and of the fundamental non-linearity of entanglement quantifiers. For such a reason the best one can do is to provide experimentally accessible bounds on some entanglement quantifiers \cite{Guhne2}. The aim of this paper is to build a bound to the entanglement degree in a general bipartition of a physical system in a mixed state. We are going to establish an upper bound to the Negativity $N$ \cite{Peres} in terms of the Linear Entropy $S_L$. We are thus studying what is called Negative Partial Transpose (NPT) entanglement. It should however be emphasized that a non-zero Negativity is a sufficient but not necessary condition to detect entanglement, since Positive Partial Transpose (PPT, or bound) entanglement exists across bipartitions of dimensions higher than $2\times 3$, which can not be detected by means of the Negativity criterion \cite{Horodecki2}. Our investigation contributes to the topical debate concerning a link between quantum correlations and mixedness \cite{Wei}. We stress  that our result is of experimental interest since the bound on $N$ may easily be evaluated measuring the Linear Entropy.

\section{An upper bound to the Negativity in terms of Linear Entropy}

Consider  a $d$-dimensional system S in a state described by the density matrix
$(0\leq p_i \leq 1, \;\;\;\forall i)$
\begin{equation}\label{rho}
\rho=\sum_ip_i{\sigma_i}
\end{equation} 
where each $\sigma_i$ represents a pure state, and define a bipartition into two subsystems S$_1$ and S$_2$ with dimensions $d_1$ and $d_2$ respectively ($d = d_1\cdot d_2$). It is common [19] to define Negativity as

\begin{equation}\label{N}
N=\frac{\|\rho^{T_1}\| -1}{d_m-1}=\frac{\textrm{Tr}{\sqrt{\rho^{T_1}(\rho^{T_1})^{\dag}}-1}}{d_m-1}
\end{equation} 
where $d_m = \min \{ d_1, d_2\}$,  $\rho^{T_1}$ is the matrix obtained through a partial transposition with respect to the subsystem $S_1$ and $ \| \cdot \| $ is the trace norm ($ \| O \|\equiv \textrm{Tr}\{ \sqrt{O O^{\dag}}\}$). In what follows we will call $d_M= \max \{ d_1, d_2\}$. By construction, $0\leq N\leq 1$, with $N=1$ for maximally entangled states only. Furthermore, the Linear Entropy $S_L$ in our system is defined as

\begin{equation}\label{SL}
S_L=\frac{d}{d-1}(1-\textrm{Tr} \rho^2)=\frac{d}{d-1}P_E
\end{equation} 
where $P_E=1-\textrm{Tr} \rho^2=1-\| \rho \|_2^2$ is a measure of mixedness in terms of the Purity $\textrm{Tr} \rho^2$ of the state, $ \| \rho \|_2 $ being the Hilbert-Schmidt norm of $\rho$  ($ \| O \|_2 \equiv \sqrt{\textrm{Tr} \{O O^{\dag}}\}$). 
By definition, $S_L = 0$ for any pure state while $S_L = 1$ for maximally mixed states.
It is easy to see that there exists a link between the trace norm of an operator $O$ in a $d$-dimensional Hilbert space and its Hilbert-Schmidt norm. Such a link can be expressed as 

\begin{equation}\label{normaO}
\| O \|^2=( \sum_{i=1}^d |\lambda_i | )^2  \leq d \sum_{i=1}^d | \lambda_i |^2 = d \| O \|^2_2
\end{equation} 
where $\lambda_i$ is the $i$-th eigenvalue of $O$ and the so-called Chebyshev sum inequality $\left( \sum_{i=1}^d a_i \right)^2  \leq d \sum_{i=1}^da_i^2$ has been used. Since, in addition, the Hilbert-Schmidt norm is invariant under partial transposition, one readily gets a first explicit link between Negativity and mixedness $P_E$, valid for generic $d$-dimensional systems, in the form of an upper bound, which reads

\begin{equation}\label{Q1}
N\leq \frac{\sqrt{d}\sqrt{1-P_E}-1}{d_m-1}\equiv Q_1
\end{equation}

Equation (\ref{Q1}) provides an upper bound to the Negativity $N$ in terms of $P_E$ and thus, in view of equation (\ref{SL}), in terms of the Linear Entropy. This bound imposes a zero value for $N$ only for a maximally mixed state. It is known \cite{Thirring}, however, that no entanglement can survive in a state whose purity is smaller than or equal to $(d-1)^{-1}$. Also in the case of a pure (or almost pure) state, the bound becomes useless as long as the bipartition is not "balanced" (by "balanced" we mean a bipartition where $d_m=\sqrt{d}$).
It indeed becomes greater than one (thus being unable to give information about entanglement) for mixedness smaller than
$\frac{d-d_m^2}{d}$  which might even approach 1 in some specific cases (recall that, by definition, $d_m\leq d$).
We however expect entanglement to be unbounded only in the case of pure states ($P_E = 0$). In the following we
show that bound (\ref{Q1}) can be strengthened.

\section{Strengthening the previous bound} Observe firstly that the rank $r_{\rho}$ of $\rho^{T_1}$ is not greater than $d_m^2$ (equal to $d$) when $\rho$ is pure (maximally mixed). For this reason we write

\begin{equation}\label{r(SL)}
r(S_L)\equiv \max_{\{\rho : \textrm{Tr} \rho^2=1-\frac{d-1}{d}S_L\}}r_{\rho}
\end{equation}

By construction, $r(0)=d_m^2$ since any pure state can be written in Schmidt decomposition consisting of $d_m$ vectors, and $r(1) = d$ because a maximally mixed state is proportional to identity. Since by definition $\left( \sum_{i=1}^d | \lambda_i |\right)^2=\left( \sum_{i=1}^{r(S_L)}{| \lambda_i |}\right)^2$ holds for any physical system, equation (\ref{Q1}) may be substituted by the following inequality

\begin{equation}\label{N_leq}
N\leq \frac{\sqrt{r(S_L)}\sqrt{1-P_E}-1}{d_m-1}
\end{equation}

Note however that there exist at least some physical systems for which the function in (\ref{r(SL)}), due to the maximization procedure involved in its definition, is always equal to $d$ in the range $S_L \in (0, 1]$, showing then a discontinuity at $S_L = 0$ as

\begin{equation}\label{SL_to_0}
\lim_{S_L\to 0} r(S_L)=d\neq d_m^2=r(0)
\end{equation}

Since we want our result to hold generally, independently on the particular system analyzed, equation (\ref{N_leq}) can not improve equation (\ref{Q1}) because even for slightly mixed states $(0 < S_L<<1) $ we have a priori no information on $r(S_L)$ which might be equal to $d$, tracing back equation (\ref{N_leq}) to (5). Despite this, we may correct (\ref{N_leq}) exploiting the expectation that for very low mixedness some of these eigenvalues are much smaller than the other ones. Indeed for all the $ r(S_L)$, not vanishing eigenvalues appearing in (\ref{normaO}) are treated on equal footing in going from $\| \rho^{T_1} \|$ to $\| \rho^{T_1} \|_2$. To properly take into account the difference between them, go back to equation (\ref{rho}) and define a reference pure state $\sigma_R$ at will among the ones having the largest occupation probability $p_R$. The spectrum of $\sigma_R^{T_1}$ consists of $n_p$ non-zero eigenvalues $\{\mu_{\alpha} ^{(R)} \}$ $\left(\max n_p =d_m^2\right)$ and of $n_m=d-n_p$ zero eigenvalues $\{ \nu_{\alpha}^{(R)}\}$.

We call the former $\alpha$-class eigenvalues and the latter $\beta$-class eigenvalues, and obviously the latter class does not contribute to $\| \sigma_R^{T_1}\|$. In order to strengthen (\ref{Q1}) we are interested in the spectrum of $\rho^{T_1}$ which, in general, consists of $d$ non-zero eigenvalues. Unfortunately, then, we can not directly introduce analogous $\alpha$- and $\beta$-classes to identify which eigenvalues contribute to the sum involved in (4) comparatively much less than the other ones, when the state $\rho$ possesses a low mixedness degree and is thus very close to a pure state.
To overcome this difficulty let us consider a parameter-dependent class of density matrices associated to the given $\rho$

\begin{equation}\label{tau}
\tau (x)=\sum_i q_i(x)\sigma_i
\end{equation} 
with $x\geq 0$, such that, for all $i$,

\begin{equation}\label{x_to_x1}
\lim_{x\to x_1} q_i(x)=p_i \;\;\;\; \lim_{x\to 0} q_i(x)=\delta_{iR}
\end{equation} 
and such that all $q_i(x)$s are continuous functions of $x$.
Thus $\tau(x)^{T_1}$ continuously connects $\rho^{T_1}$ and $\sigma_R^{T1} $ and, as a consequence, any $\nu_{\beta}^{(R)}$ is continuously connected to a particular eigenvalue of $\rho^{T_1}$ which will be the the corresponding mixed state $\beta$-class eigenvalue $\nu_{\beta}$. In such a way one can define the function $\nu_{\beta_0}(x)$ as the eigenvalue of $\tau(x)^{T_1}$ having the property 

\begin{equation}\label{x_to_0}
\lim_{x\to 0} \nu_{\beta_0}(x)=\nu_{\beta_0}^{(R)}
\end{equation}
and so the $\beta$-class eigenvalue for $\rho^{T_1}$ as
\begin{equation}\label{nu_beta}
\nu_{\beta_0}\equiv \lim_{x\to x_1} \nu_{\beta_0}(x)
\end{equation}

We emphasize at this point that the results of this paper do not depend on the explicit functional dependence of $\tau(x)$ on $x$, which can be chosen at will provided it satisfies conditions (\ref{x_to_x1}). Indeed, $\tau(x)$ is just a mathematical tool, with (in general) no physical meaning.
To save some writing and in view of equation (\ref{normaO}), we put 
\begin{equation}\label{A}
A=\sum_{\alpha}\mu_{\alpha}^2 \;\;\;\; B=\sum_{\beta}\nu_{\beta}^2 
\end{equation} 
and notice that $\textrm{Tr}(\rho^{T_1})^2=\textrm{Tr} \rho^2=A+B$.
We can now state (see Appendix for a proof) the following.

{\bf Lemma 1} Given a state $\rho$ of a system in a $d$-dimensional Hilbert space, and the associated reference pure state $\sigma_R$, for any set of states $\tau(t)$ satisfying (\ref{tau}) and (\ref{x_to_x1}), there exists a value $\delta\geq x_1$ such that $1-A(t)-B(t)\geq B(t)d$ for any $t \in [0,\delta]$.
This result allows us to find a function $w(S_L)$ such that $w(0) = d_m^2$ and

\begin{equation}\label{normarho}
\| \rho^{T_1} \|^2\leq w(S_L)=f(\|\rho^{T_1} \|^2_2)
\end{equation}

Starting from the identity
\begin{eqnarray}\label{identity}
\| \rho^{T_1} \|^2=(\sum_{\alpha}^{d_m^2}|\mu_{\alpha}|)^2+(\sum_{\beta}^{d-1}|\nu_{\beta}|)^2+
 \sum_{\alpha}^{d_m^2}|\mu_{\alpha}| \sum_{\beta}^{d-1}|\nu_{\beta}|
\end{eqnarray}
and applying the Chebyshev sum inequality term by term we obtain

\begin{equation}\label{rho_T1}
\| \rho^{T_1} \|^2\leq ( d_m\sqrt{A+B}+\sqrt{\frac{d-1}{d}}\sqrt{1-A-B} )^2
\end{equation}
where Lemma 1 has been exploited. Expressing equation (\ref{rho_T1}) in terms of Negativity and Purity we finally get

\begin{equation}\label{Q2}
N\leq \frac{d_m\sqrt{1-P_E}+\sqrt{\frac{d-1}{d}}\sqrt{P_E}-1}{d_m-1}\equiv Q_2
\end{equation}

Bound (\ref{Q2}) improves bound (\ref{N_leq}) for high purity when $S_L$ is small, that is $Q_2 < Q_1$, becoming in general greater than $Q_1$ at low purity. In addition it still suffers the same drawback as $Q_1$, not vanishing when $1-P_E=\frac{1}{d-1}$. 
In such a case one has to consider the lower bound $\frac{1}{d-1}$ on purity, below which no entanglement survives. In order to take such a bound into account, instead of distinguishing among $\alpha$ and $\beta$ eigenvalues of $\rho^{T_1}$ , we can divide them into positive ones $\{ \xi_i\}$ and negative ones $ \{\chi_i\}$. In this way, calling $n_-$ and $ (d-n_-)$ the numbers of negative and positive eigenvalues, respectively and applying the Lagrange multiplier method to the function $\| \rho^{T_1} \|=\sum_i^{d-n_-} \xi_i+\sum_j^{n_-} \chi_j$ subjected
to constraints $\sum_i\xi_i^2+\sum_j \chi_j^2=1-P_E$ and $\sum_i\xi_i+\sum_j \chi_j=1$ one finds

\begin{equation}\label{norma_rho_leq}
\| \rho^{T_1} \|^2\leq\frac{d-2n_-+2\sqrt{n_-(d-n_-)(d(1-P_E)-1)}}{d}
\end{equation}

Bound (\ref{norma_rho_leq}) can be exploited to show that no entanglement can survive at purity lower than $\frac{1}{d-1}$. Indeed, for entanglement to exist, one eigenvalue at least has to be
negative. However, by normalization, it always has to be true that $\| \rho^{T_1} \| \geq 1$ and this implies that, as long as $n_-\geq 1$, purity $1-P_E$ can not be smaller than $\frac{1}{d-1}$ as expected. However, in general, the number of negative eigenvalues is not known. In these cases the best one can do is to look for the maximum, with respect to $n_-$, of the right-hand side of (\ref{norma_rho_leq}), leading unfortunately once again to bound (\ref{Q1}) on $N$. However, since always $N\leq \Theta (\frac{d-2}{d-1}-P_E)\equiv Q_3$, where $\Theta (x)$ is the Heaviside step function, defining $Q=\min\{Q_1, Q_2, Q_3\}$, we state our final  and main result as

\begin{equation}\label{NQ}
N \leq Q
\end{equation} 
valid for every possible bipartition of a quantum system, independently on its (finite) dimension, its detailed structure or its properties. It is worth stressing that computing Negativity quickly becomes a very hard task as the dimension of the Hilbert space grows, while the evaluation of purity can be performed without particular efforts. We emphasize in addition that bound $Q$ in (\ref{NQ}) only depends on purity, and is completely  determined once a bipartition of the physical system is fixed and purity is known. This
means that an experimental measure of purity allows to extract information about the maximal degree of bipartite entanglement one can find in the system under scrutiny.
Some purity measuring protocols, or at least purity estimations based on experimental data, have been proposed. They are based on statistical analysis of homodyne distributions, obtained measuring radiation field tomograms\cite{Manko}, on the properties of graph states \cite{Wunderlich}, or on the availability of many different copies of the state over which separable measurements are performed \cite{Bagan}. In all the cases where a measure of purity is possible, an experimental estimation of bipartite entanglement is available thanks to
(\ref{NQ}) which is then actually experimentally accessible.

\section{Crossover between $Q_1$ and $Q_2$ and numerical results} 
As commented previously, bounds $Q_1$ (Eq. (\ref{Q1})) and $Q_2$ (Eq. (\ref{Q2})) can supply information about bipartite entanglement in two different setups: $Q_1$ is indeed accurate enough for a balanced bipartition (i.e. when $d_m\sim \sqrt{d}$) but fails when $\sqrt{d}\gg  d_m$ since it rapidly becomes greater than 1. To solve this problem, we obtained the bound $Q_2$ which, by construction, provides nontrivial information about bipartite entanglement in an unbalanced bipartition $(\sqrt{d}\gg d_m )$, but may not work properly for a
balanced one. It is actually a very easy task to show that our new bound $Q_2$ works better than the old one,
$Q_1$, (i.e., $Q_1\geq Q_2$) when the purity $P$ is greater than a critical value given in terms of the total Hilbert space dimension and of the subdimensions of the bipartition, i.e. when

\begin{equation}\label{Pc}
P(\rho)\geq \frac{d-1}{d(\sqrt{d}-d_m)^2+d-1}=P_c
\end{equation}

\begin{figure}
\includegraphics[width=0.60\textwidth, angle=0]{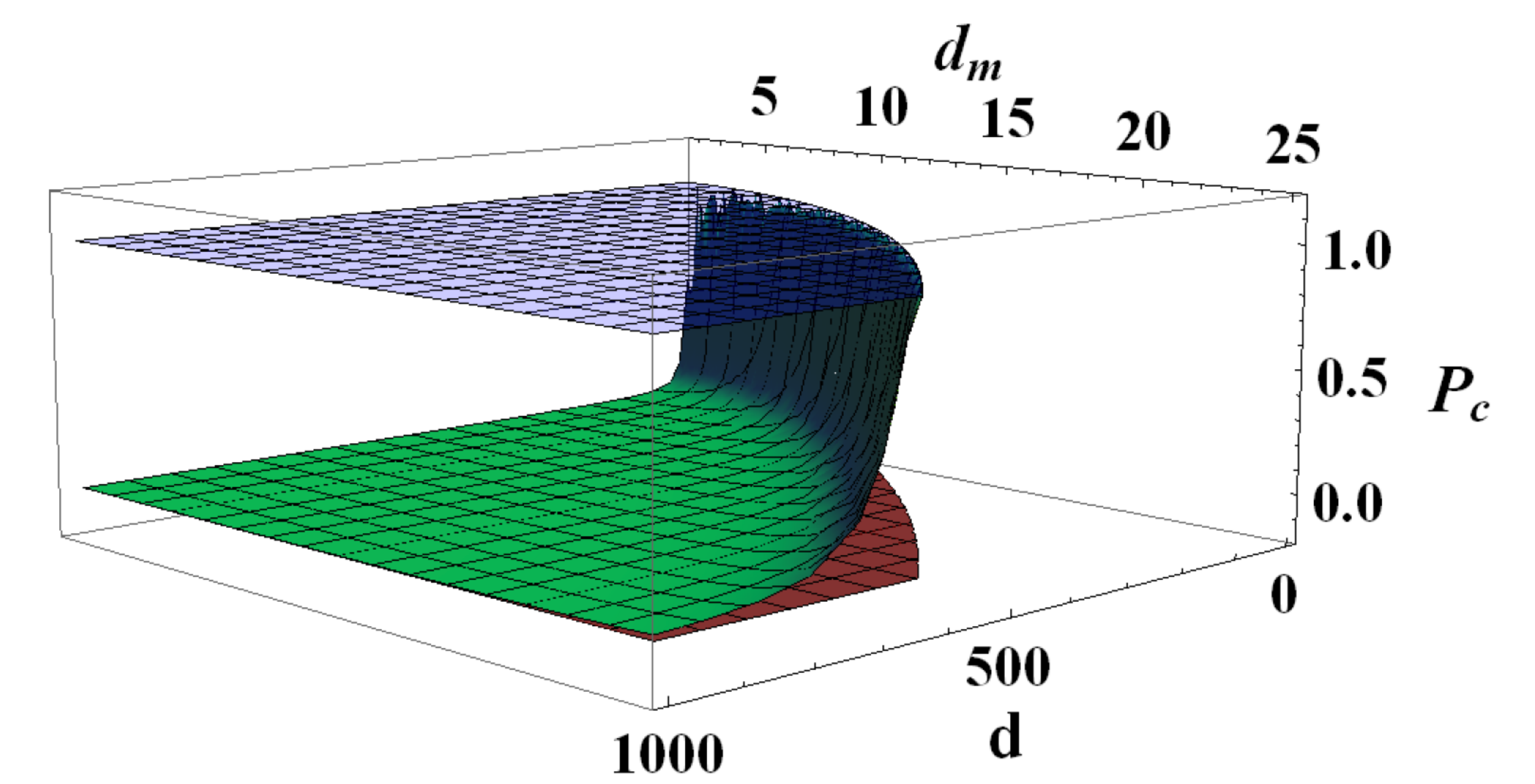}
\caption{(Color online) Purity threshold $P_c$ given in Eq. (\ref{Pc}) (green surface) and natural purity limits 1 (blue upper surface) and $\frac{1}{d}$ (red lower surface) as functions of $d\in [4,1000]$ and $d_m\in [2,25]$ such that $d_m^2\leq d$. Values of purity below the green surface are such that $Q_1 < Q_2$, while values of purity above the green surface yield $Q_2 < Q_1$. It is clear that, when $d_m^2< d$, $P_c\sim \frac{1}{d}$ meaning that $Q_2 < Q_1$ for most of quantum states. On the other hand, when $d_m^2\sim d$ and $Q_1 < Q_2$ almost everywhere in state space. } 
\label{Fig1}
\end{figure}

Two limiting cases are easily studied directly from Eq. (\ref{Pc}): for a perfectly balanced bipartition ($\sqrt{d}=  d_m$), one gets $P_c=1$ and, since by definition $\frac{1}{d}\leq P(\rho)\leq 1$, in this case the bound $Q_1$ is smaller (and therefore works better) than the bound $Q_2$ for any possible quantum state. In the opposite limit, in a strongly unbalanced bipartition, one can roughly approximate $(\sqrt{d}-d_m)^2\sim d$ and,
since by definition $d\geq 4$, this leads to $P_c\sim \frac{d-1}{d^2+d-1} < \frac{1}{d}$. Taking into account the natural bounds for the purity of a quantum state, this in turn means that in such a limit $Q_1 >  Q_2$ for any quantum state or, in other words, our new bound works always better.
This behavior can be clearly seen in Fig. (\ref{Fig1}), where the dependence of $P_c$ on $d$ and $d_m$ is shown together with the natural limiting values of $P(\rho)$.

\begin{figure}
\includegraphics[width=0.60\textwidth, angle=0]{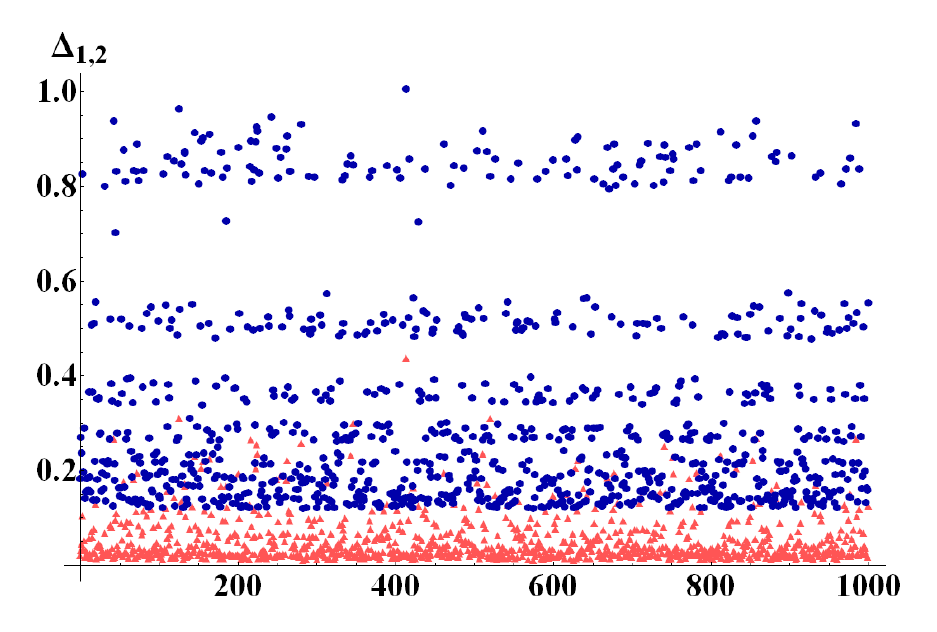}
\caption{((Color online) $\Delta_1$ (light red triangles) and $\Delta_2$ (dark blue circles) evaluated for 1000 randomly generated bipartite states, with $d_m = d_M$ randomly chosen in [2, 10]. For these perfectly balanced bipartite states $\Delta_1 < \Delta_2$ everywhere in state space. } \label{Fig2} 
\end{figure}

\begin{figure}
\includegraphics[width=0.60\textwidth, angle=0]{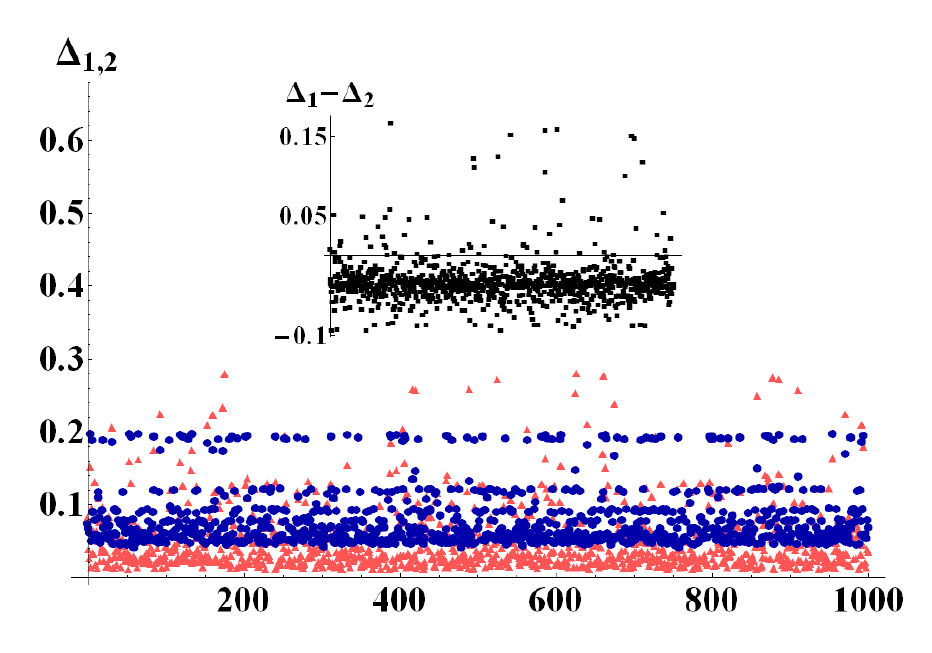}
\caption{ (Color online) $\Delta_1$ (light red triangles) and $\Delta_2$ (dark
blue circles) evaluated for 1000 randomly generated bipartite states, with $d_M = d_m+60$ and $d_m$ randomly chosen in [2, 10]. Since the bipartitions are no longer perfectly balanced, there is a much broader mixing of values of $\Delta_1$ and $\Delta_2$. In particular, $\Delta_1$ has a much wider distribution of values, while $\Delta_2$ seems to have a much denser distribution around central values. The inset shows the difference $\Delta_1-\Delta_2$. }
\label{Fig3}
\end{figure}

\begin{figure}
\centering
\includegraphics[width=0.60\textwidth, angle=0]{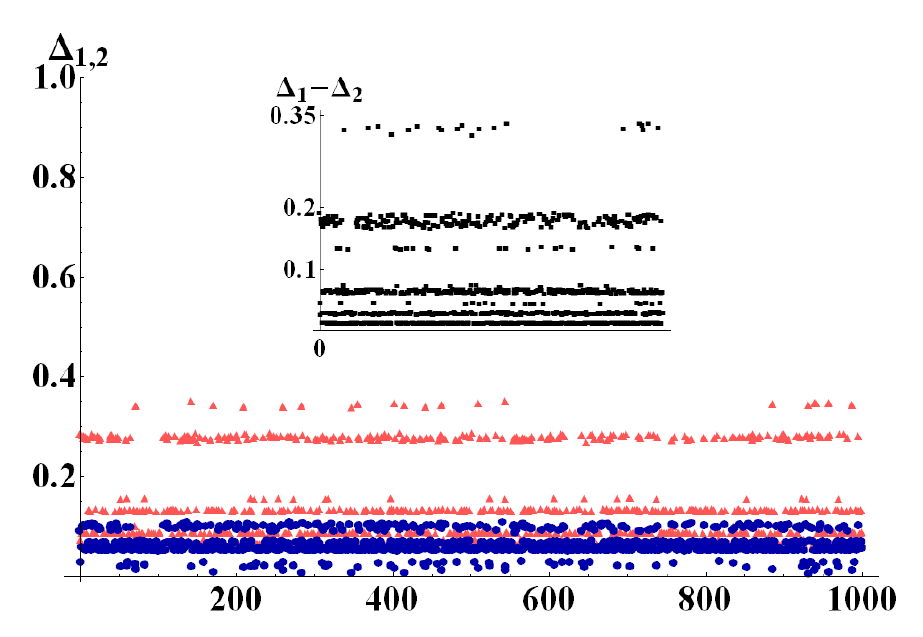}
\caption{(Color online) $\Delta_1$ (light red triangles) and $\Delta_2$(dark blue circles) evaluated for 1000 randomly generated bipartite states, with  $d_M = d_m+70$ and $d_m$ randomly chosen in [2, 5]. For these strongly unbalanced bipartitions we always detect $\Delta_1>\Delta_2$. The inset shows the difference $\Delta_1-\Delta_2$. } 
\label{Fig4}
\end{figure}

To better exemplify such a behavior, we report here results of numerical simulations performed with the aid of the $QI$ package for Mathematica \cite{Mathematica}, by which random quantum states have been generated in different dimensions, uniformly distributed according to different metrics. On these states, we tested bounds $Q_1$ and $Q_2$.
Figures (\ref{Fig2})-(\ref{Fig4}) show the differences $\Delta_i=Q_i-N$ $(i=1,2)$ of the bounds with the Negativity of the state, once a bipartition has been fixed. In particular, in a first run of simulations (Fig. (\ref{Fig2})) we generated $10^3$ perfectly balanced bipartite states (such that $d_m=d_M=\sqrt{d}$), randomly choosing the dimension of the two subsystems for each quantum state within the range $d_M=d_m\in[2,10]$. The results in Fig. (\ref{Fig2}) clearly show that $\Delta_1< \Delta_2$ for all the analyzed states.
The second run of simulations has been performed with $d_m$ randomly chosen in $[2, 14]$ and $d_M = d_m+60$. In such a case, as can be seen in Fig. (\ref{Fig3}) the difference $\Delta_1-\Delta_2$ has no fixed sign. The two subdimensions are, indeed, such that the critical value of purity $P_c$ in Eq. (\ref{Pc}) is neither
extremely close to $\frac{1}{d}$ nor to 1. As can be noticed from the inset of Fig. (\ref{Fig3}) which shows the difference $\Delta_1-\Delta_2$, however, on the average it is still true that $\Delta_1<\Delta_2$.
The third set of numerical data, finally, has been obtained generating $10^3$ random states with subdimensions $d_M=d_m+70$ and $d_m$ randomly drawn in $[2,5]$. In this limit the value of $P_c$ is very close to the minimum of purity and we therefore expect $Q_2$ to work better than $Q_1$ for almost any state. This is indeed confirmed by the simulations shown in Fig. (\ref{Fig4}), in which $\Delta_2<\Delta_1$.
As an example of application of our results, consider a single two-level system interacting with a spin system composed of $n_s$ spins, each of which lives in a $d_s$ dimensional Hilbert space. Therefore the total system Hilbert space has dimension $d = 2(d_s)^{n_s}$ . Let us suppose the spin system is a chain of 10 spin $\frac{1}{2}$ (which is a relatively small system, very far from its thermodynamic limit). The total Hilbert space dimension will then be $d = 2^{11}$, and considering the natural bipartition into the two-level system and the spin chain, one has $d_m = 2$ and $d_M = 2^{10}$. For such a system, the critical value of purity $P_c$ in Eq. (\ref{Pc}) is

\begin{equation}\label{Pcnum}
P_c=\frac{2^{11}-1}{2^{11}(\sqrt{2^{11}}-2)^2+2^{11}-1}\sim 0.000534
\end{equation}

The lower value of purity for which bipartite entanglement can survive is, as said before, $P_l=\frac{1}{d-1}\sim 0.00049$. Therefore, for all the total states having purity $0.000534\leq P(\rho)\leq 1$, bound $Q_2$ in Eq. (\ref{Q2}) works better than $Q_1$. Only for the small fraction of states having an extremely low purity in the range $[0.00049, 0.000534]$ bound $Q_1$ gives better information than $Q_2$. This shows again that, for unbalanced bipartitions (and even in the case of a relatively small number of individual components of the total system), $Q_2$ works much better than $Q_1$.

\section{ Application to thermal entanglement} Of particular interest is the application of the results of this paper to the case of thermal entanglement, where both Linear Entropy and its link to Negativity acquire a much clearer meaning. A recent result \cite{Popescu}, indeed, shows how the canonical ensemble description of thermal equilibrium stems from the existence of quantum correlations between a system and its thermal bath. In view of this it has been shown that it is possible, with a very small statistical error, to replace the system + bath microcanonical ensemble with a pure state inside the suitable energy shell, still obtaining the appropriate thermal statistics characterizing Gibbs distribution. In this context, then, the Linear Entropy of the mixed Gibbs state provides a system/bath
entanglement measure. Equation (\ref{NQ}) then can be viewed as a monogamy relation, describing the competition between two kinds of quantum correlations - internal ones measured by Negativity and external ones measured by Entropy.
On the other hand it is known that some thermodynamic quantities (like e.g. heat capacity or internal energy) can be used as entanglement witnesses \cite{Wiesniak}, and recent works have shown an even closer link between heat capacity and entanglement for particular systems \cite{Leggio1,Leggio2}. The result of this paper suggests this link might hold very generally. Indeed, in the case of a Gibbs equilibrium state, $P_E$ can be given by the expression

\begin{equation}\label{PE}
P_E=\sum_{i\neq j}\frac{e^{-\beta E_i}e^{-\beta E_j}}{Z^2}=\sum_{i\neq j}P_{E}^{ij}
\end{equation} 
where $E_i$ is the $i$-th energy level of the system and $Z$ is its partition function, $\beta$ being the inverse temperature in units of $k_B$. Heat capacity in a finite dimensional system reads

\begin{equation}\label{CV}
C_V\equiv\beta^2(\langle H^2\rangle-\langle H\rangle^2)=\beta^2\sum_{i\neq j}P_E^{ij}\frac{E_i-E_j}{2}
\end{equation}

There is then a similarity between $P_E$ and $C_V$ as given by equations (\ref{PE}) and (\ref{CV}), suggesting how a measure of the latter, together with little knowledge about the energy spectrum of the physical system, might supply significant information on the Linear Entropy of the system and, as a consequence, on its maximal degree of internal bipartite entanglement. This triggers interest in further future investigation on a detailed analysis of the relation between  $P_E$ and $C_V$ which, in turn, might supply us with an easily experimentally measurable entanglement bound as well as highlight how the origin of thermodynamical properties is strongly related to non-classical correlations and monogamy effects. Such a connection, and the usefulness of the bounds derived in the previous sections, can be exemplified with a simple three qutrit system with a parameter-dependent Hamiltonian

\begin{equation}\label{Hl}
H_l=\omega J_z+\tau {\bf J}_1\cdot {\bf J}_2+({\bf J}_1\cdot {\bf J}_2)^2+k{\bf J}_0\cdot({\bf J}_1+{\bf J}_2)
\end{equation} 
where ${\bf J}_i$ is the spin operator of the $i$-th particle, ${\bf J} ={\bf J}_0 +{\bf J}_1 +{\bf J}_2$.  and $\omega,\;\; \tau, \;\;k$ are real interaction parameters. This effective Hamiltonian operator describes a system consisting of two ultracold atoms (spins labeled as 1 and 2) in a two-well optical lattice and in the Mott insulator phase, where thus the tunneling term in the usual Bose-Hubbard picture is accounted for as a second order perturbative term, both coupled with a third atom (labeled as 0) via an Heisenberg-like interaction. An external magnetic  field is also present, uniformly coupled to the three atoms. Such a system is a generalization of the one studied in \cite{Leggio2}, where a deep connection between thermal entanglement and heat capacity in parameter space has been shown. Hamiltonian (\ref{Hl}) is analytically diagonalizable, thus allowing us to obtain explicit expressions for thermodynamic quantities characterizing the Gibbs equilibrium state of the three-atom system, together with the Negativity of the reduced state of the two quadratically coupled spins. 

\begin{figure}
\centering
\includegraphics[width=0.60\textwidth, angle=0]{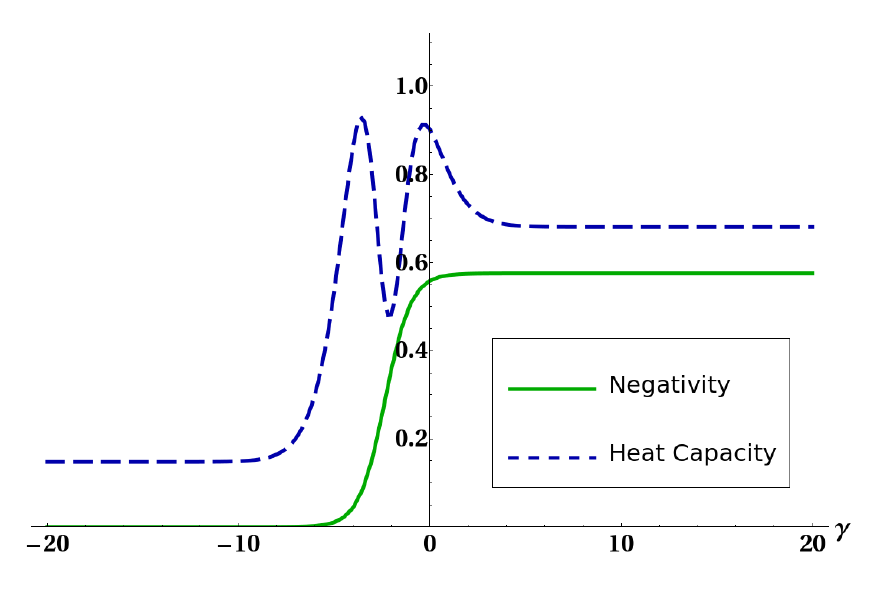}
\caption{Negativity of reduced state of the two ultracold atoms (full line) and Heat Capacity of the system (dashed line) versus quadratic interaction parameter $\gamma$. The other parameters have been fixed as $k_B T=2$, $\tau=3$,  and $k=1$. } 
\label{Fig5}
\end{figure}

\begin{figure}
\centering
\includegraphics[width=0.60\textwidth, angle=0]{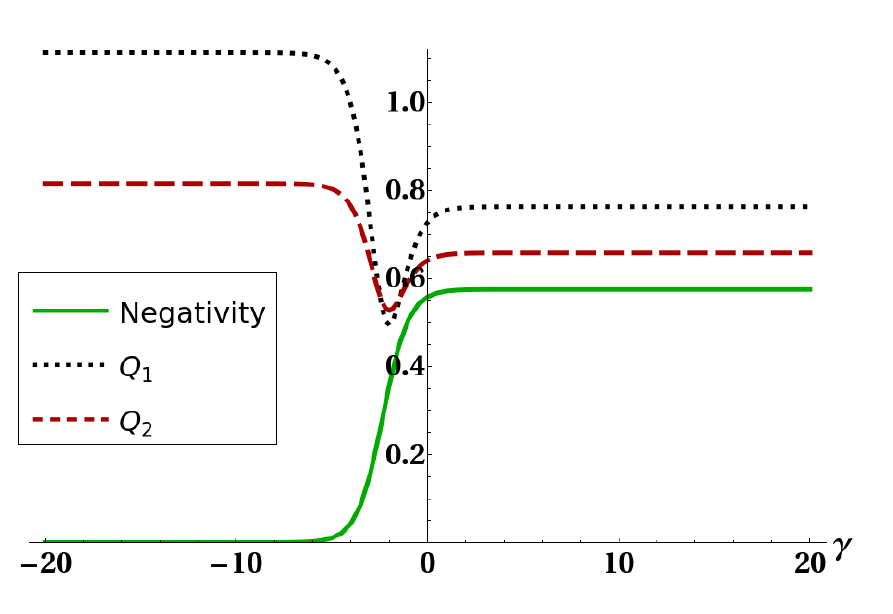}
\caption{Negativity of reduced state of the two ultracold atoms (full line), bound $Q_1$ (dotted line) and bound $Q_2$ (dashed line) versus quadratic interaction parameter $\gamma$. The other parameters have been fixed as $k_B T=2$, $\tau=3$,  and $k=1$. }
\label{Fig6}
\end{figure}

\begin{figure}
\centering
\includegraphics[width=0.60\textwidth, angle=0]{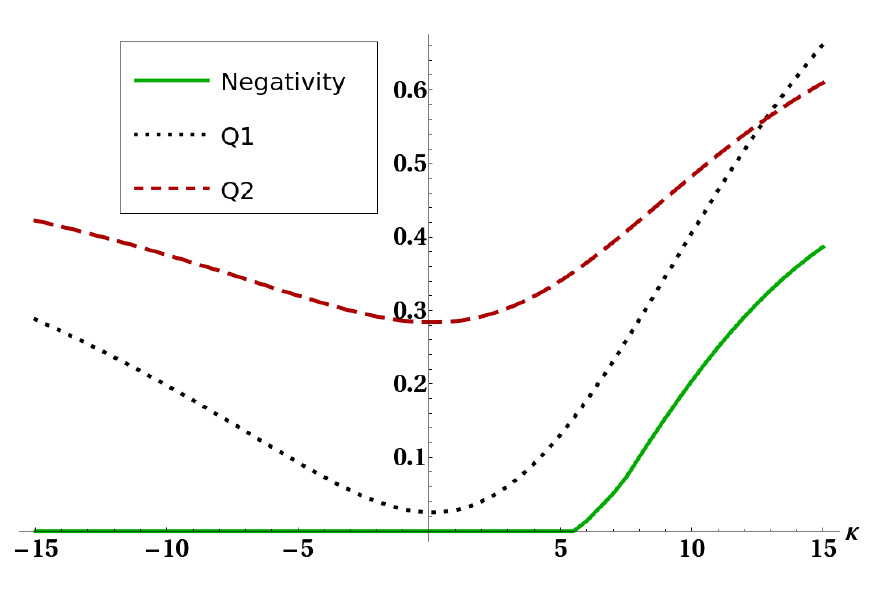}
\caption{Negativity of reduced state of the two ultracold atoms (full line), bound $Q_1$ (dotted line) and bound $Q_2$ (dashed line) versus Heisenberg interaction parameter $k$. The other parameters have been fixed as $k_B T=10$, $\tau=3$,  and $\gamma=1$.  } 
\label{Fig7}
\end{figure}

\begin{figure}
\centering
\includegraphics[width=0.60\textwidth, angle=0]{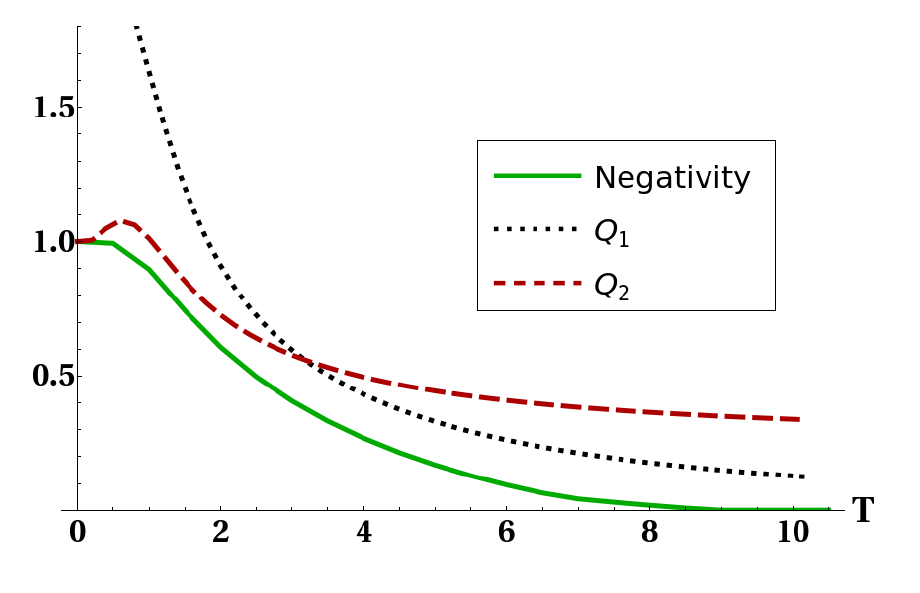}
\caption{Negativity of reduced state of the two ultracold atoms (full line), bound $Q_1$ (dotted line) and bound $Q_2$ (dashed line) versus temperature $T$ (in units of $k_B$). The interaction parameters have been fixed as $\tau=4$, $k=5$ and $\gamma=1$. } 
\label{Fig8}
\end{figure}

The mathematical origin of the connection between heat capacity and Negativity was already discussed in  \cite{Leggio2} and is ultimately due to the presence of level crossing in the low-lying energy eigenvalues of the system. Here we want to show how the existence of the strong connection between purity and Negativity, expressed by bound (\ref{NQ}), can give some hints for a physical explanation of such an effect, and moreover to exemplify how bound (\ref{NQ}) can often supply important information on the amount of thermal entanglement. Figure (\ref{Fig5}) shows how the connection between thermal entanglement and  heat capacity highlighted in  \cite{Leggio2} is still present despite the interaction with a third atom. Figures (\ref{Fig6}) and (\ref{Fig7}) show bounds (\ref{Q1}) and (\ref{Q2}), together with the Negativity of the reduced state of two atoms, versus a certain interaction parameter in the Hamiltonian. Figure (\ref{Fig8})  finally shows the same quantities versus temperature for fixed Hamiltonian parameters. All energies in the plots are expressed in units of $\omega$. It is worth stressing here that, in all these plots, bound $Q_3=\Theta(\frac{d-2}{d-1}-P_E)$ is not shown. The reason is that, in order to preserve thermal entanglement, temperature in our simulations has to
be kept at most of the same order of magnitude of spin-spin interactions, and in such a regime $P_E$ has not yet crossed the threshold $\frac{d-2}{d-1}$  so that $Q_3$ is constantly equal to one.
It is clearly shown in Figures (\ref{Fig6}) and (\ref{Fig8}) how bound $Q_1$ given in (\ref{Q1}) can become, as discussed, larger than 1. In all these cases (except for a small temperature range in Figure (\ref{Fig8}), however, (\ref{Q2}) is still able to sensibly bound Negativity. In all the plots shown, and in general every time the bounds (\ref{Q1}) and (\ref{Q2}) are applied to the particular system analyzed here, one always gets useful information about bipartite entanglement in the form, of course, of an upper bound. Such a bound, however, gets very close to zero in some particular cases (see, for example, Figures (\ref{Fig7}) and (\ref{Fig8})), strongly restricting the allowed range of values for the Negativity. It is then shown that (\ref{NQ}) is able to produce non-trivial results. 
 It is worthy noting from that in Fig.  (\ref{Fig5}) there exist ranges of the parameter  $\gamma$ where the Negativity and the Heat Capacity  exhibit simultaneous  plateaus. This fact, previously shown and commented in reference \cite{Leggio2} too, in view of Equation (\ref{NQ}) and the strong link between Heat Capacity and the mixedness $P_E$ of a quantum state legitimates the deduction that in the parameter  regions of very low Negativity the Heat capacity may be assumed as almost constant. 


\section{Conclusions}
In this paper we derived a bound on the degree of information storable as bipartite quantum entanglement within an open $d$-dimensional quantum system in terms of its Linear Entropy. Our result is quite general, holding for arbitrary bipartitions of an as well arbitrary system.  We emphasize that our result is experimentally appreciable in view of quite recently proposed protocols aimed at measuring the Purity of a state of a quantum system.  Inspired by the seminal paper of Popescu, Short and Winter \cite{Popescu}, our conclusions highlight the interplay between quantum entanglement inside a thermalized system and its physical properties.  Our results are of interest not only for the Quantum Information researchers but also for the growing cross-community of theoreticians and experimentalists investigating
the subtle underlying link between quantum features and thermodynamics.

%
%
\section{Appendix}
{\bf Proof of Lemma 1.} Let us first notice that both $B(t)$ and $1-A(t)-B(t)$ go to zero when the state $\tau(t)$ is pure. Indeed $A+B=\textrm{Tr} \tau^2$ equals the
purity, while $B$ by construction vanishes when $\tau$is pure (see Equations (9)-(13)). Notice further that both $A(t)$ and $B(t)$ are quadratic in $\nu_{\beta}(t)$ and $\mu_{\alpha}(t)$  and then the statement of Lemma 1 is independent of their sign. Such a statement can be rewritten as

\begin{equation}\label{A1}
\frac{1-A-B}{d}-B=l(A, B)\geq 0
\end{equation}

The function $l(A,B)$ at the extremal points of its domain (corresponding to a pure and a maximally mixed state) satisfies (\ref{A1}). For a maximally mixed state, calling $n_{\alpha}$ $(n_{\beta})$ the number of $\alpha$- $(\beta-)$class eigenvalues $(n_{\alpha}+n_{\beta})$, one gets $A=\frac{n_{\alpha}}{d^2}$ and $B=\frac{n_{\beta}}{d^2}$ and thus $l=\frac{n_{\alpha-1}}{d^2}\geq 0$ since $n_{\alpha}\geq 1$.
Let us now express $l(A,B)$ as

\begin{equation}\label{A2}
h(\{\mu_{\alpha}\}, \{\nu_{\beta}\})=\frac{1}{d}\left( 1-\sum_{\alpha}^{n_{\alpha}}\mu_{\alpha}^2- \sum_{\beta}^{n_{\beta}}\nu_{\beta}^2\right)- \sum_{\beta}^{n_{\beta}}\nu_{\beta}^2
\end{equation}

We can address the investigation on internal points using the Lagrange multiplier method, taking into account the trace condition $\sum_{\alpha}^{n_{\alpha}}\mu_{\alpha}^2+ \sum_{\beta}^{n_{\beta}}\nu_{\beta}^2= 1$. 
From this method only one stationary point results, characterized by values of $\nu_{\beta}$ and $\mu_{\alpha}$ such that the corresponding state is mixed. It is straightforward to check that at this point the function (\ref{A2}) is positive. We then deduce that $l(A,B)\geq 0$, from which Lemma 1 directly follows. Finally, the range $\delta$ of validity of Lemma 1 is given by the requirement $q_R(t)\geq q_{i\neq R}(t)$, such a property being necessary for the sensible definition of the reference pure state $\sigma_R$ which guarantees, in turn, that $B(t)$ vanishes on pure states.
%

%
\vspace{6pt} 





\acknowledgments{ A.M. acknowledges the kind hospitality provided by HN at the Physics Department of Waseda University on November 2019. All the authors are grateful to Marek Kus for his constructive and stimulating suggestion and for carefully reading the manuscript. HN was partly supported by Waseda University Grant for Special Research Projects (Project number: 2019C-256).}

\end{document}